\documentclass[11pt,superscriptaddress,aps,prd,preprint]{revtex4}

\usepackage[utf8]{inputenc} 
\usepackage{amsmath}
\usepackage{mathrsfs}
\usepackage{graphicx}
\usepackage{amssymb, theorem, xspace}
\usepackage{pgfplots}
\usepackage{dcolumn}
\usepackage{slashed}
\usepackage{xcolor}

\newcommand{\newc}{\newcommand}
\newc{\del}{\partial}
\newc{\rpdv}[1]{\frac{\overrightarrow{\partial}}{\partial{#1}}}
\newc{\lpdv}[1]{\frac{\overleftarrow{\partial}}{\partial{#1}}}
\newc{\at}[2][]{#1\bigg|_{#2}}
\newc{\bea}{\begin{eqnarray}}
\newc{\eea}{\end{eqnarray}}

\begin{document}

\title{Size-effect  at finite temperature in a quark-antiquark effective model in phase space }

\author{G. X. A. Petronilo}
\email{gustavo.petronilo@aluno.unb.br}
\affiliation{Instituto de F\'{i}sica, Universidade de Bras\'{i}lia, 70910-900, Bras\'{i}lia, DF, Brazil.}

\author{R. G. G. Amorim}
\email{ronniamorim@gmail.com }
\affiliation{Faculdade Gama, Universidade de Bras\'ilia, Setor Leste (Gama), 72444-240 Bras\'ilia, DF, Brazil} \affiliation{Canadian Quantum Research Center,\\
204-3002 32 Ave Vernon, BC V1T 2L7  Canada}

\author{S. C. Ulhoa }
\email{sc.ulhoa@gmail.com}
\affiliation{Instituto de F\'{i}sica, Universidade de Bras\'{i}lia, 70910-900, Bras\'{i}lia, DF, Brazil.} \affiliation{Canadian Quantum Research Center,\\
204-3002 32 Ave Vernon, BC V1T 2L7  Canada}

\author{A. F. Santos}
\email{alesandroferreira@fisica.ufmt.br}
\affiliation{Instituto de F\'{\i}sica, Universidade Federal de Mato Grosso,\\
78060-900, Cuiab\'{a}, Mato Grosso, Brazil.}

\author{A. E. Santana}
\email{a.berti.santana@gmail.com }
\affiliation{Instituto de F\'{i}sica, Universidade de Bras\'{i}lia, 70910-900, Bras\'{i}lia, DF, Brazil.}

\author{Faqir C. Khanna\footnote{Professor Emeritus - Physics Department, Theoretical Physics Institute, University of Alberta - Canada}}
\email{fkhanna@ualberta.ca}
\affiliation{Department of Physics and Astronomy, University of Victoria,
3800 Finnerty Road Victoria, BC, Canada.}


\begin{abstract}

A  quark-antiquark effective model is studied in a toroidal topology at finite temperature. The model is described by a Schr\"odinger equation with linear potential which is embedded in a torus. The following aspects are analysed: (i) the nonclassicality structure using the Wigner function formalism; (ii) finite temperature and size-effects are studied by a generalization of Thermofield Dynamics written in phase space; (iii) in order to include the spin of the quark, Pauli-like Schr\"odinger equation is used; (iv) analysis of the size-effect is considered to observe the fluctuation in the ground state. The size effect goes to zero at zero, finite and high temperatures. The results emphasize that the spin is a central aspect for this quark-antiquark effective model.

\end{abstract}



\maketitle
\section{Introduction}

The present work is motivated by the successful approach describing a sector of Quantum Chromodynamics (QCD) where the quark-antiquark interaction is the so-called funnel potential, increasing linearly with the distance~\cite{godfrey,brambilla,martin2,lich2}. In this sector,  hadrons are bound states of quarks that are described by a non-relativistic Galilean potential. The time evolution of the field is carried by the Schr\"odinger equation.

Since hadrons are confined states of QCD, three aspects deserve a detailed analysis. The first is the non-classicality of a quantum state, which is studied using the Wigner formalism of the phase space. Another  aspect is an investigation of the spin 1/2 nature of quark. The third element is an analysis of the confinement at a finite temperature. These two later aspects can be improved by considering the size-effect of the vacuum fluctuation at finite temperature.  Regarding the vacuum fluctuation, some analyses of the so called Casimir effect are only partially explored~\cite{Our0} and by considering the non-relativistic limit of the relativistic massive Klein-Gordon field, the results are inconclusive~\cite{cougo2001}. Our main goal is to address these three parts, by considering the symplectic quantum field theory formulated in a torus at finite temperature. For that a symplectic version of thermofield dynamics is used. To assure the Galilei symmetry of the space-time, throughout the calculations,  the formalism is written in the light-cone of the  $(4+1)$-de Sitter space~\cite{Pinski, Duval, Takahashi, humi, Santana}.

Galilei symmetry in the classical mechanics, despite all the time that has elapsed and progress achieved, is not exhausted. Some phenomenon
remain challenging~\cite{size2,size3,size4,size5}. On the other hand, it is well-known that tensor form for the quantum field theory and general relativity are important. Naturally the question arises: is it possible to write Galilei symmetry in a tensorial form, using linear transformations? The answer is affirmative and known as the Galilean covariance. This creates a geometric structure analogous to the Minkowski space-time~\cite{Duval, Takahashi,Santana}. However without imposing a limit for information, the Galilean covariance is achieved through an immersion of the Euclidean space, $R^3$, into a $(4+1)$-de Sitter space, $G$. With this formulation it is possible to write the Schr\"odinger equation in a form similar to Klein-Gordon equation in $(4+1)$-de Sitter space. The Galilean covariance also enables a study of a version of the Dirac equation in non-relativistic approach. This is the Pauli-Schr\"odinger equation.

The covariant version of non-relativistic equations has made it possible to analyse problems that do not have solutions known in the usual formalism, such as the study of spin for low velocity systems. The Galilean covariance allows breaking  of the spin degeneration without the use of perturbation theory. Such effects have been investigated already in Thermo Field Dynamics (TFD)~\cite{Our0}, that introduces temperature effects. However, an analysis of such a system in phase space is still missing. This is of interest since such an investigation sheds light on the statistical nature of quantum states in compactified conditions.

The Galilean covariant tensor formalism in phase space using the approach of symplectic quantum mechanics has been developed recently~\cite{petron23, Amorim2020}. The path integral formalism of Galilean covariance in phase space is developed to consider the calculations of TFD. The calculations are carried out in phase space of relativistic fields~\cite{Our, Our1}.
A main goal is to explore the Galilean covariance in phase space and to investigate the size-effect of non-relativistic quantum systems compactified in a torus. Considering that the quark-antiquark interaction is phenomenologically described by a linear (Galilean) potential, the spin 1/2 of quarks is included, by using a Galilei-covariant Pauli-Schr\"odinger equation in phase space. Then the size-effects are analysed in the vacuum fluctuation using the TFD formalism  to deal with the spatial torus at finite temperature with Wigner functions.

The formalism of quantum mechanics in phase space was introduced by Wigner \cite{Wigner,Hillery}. This allows mapping of classical functions into quantum operators in phase space using the Moyal product, such that Wigner functions are mapped in a quasi-probability density using the Hilbert space Euclidean wave functions. Another step in the quantum phase space is carried out in order to accommodate  consistently the notion of gauge field with the Wigner function. This is accomplished with the notion of quasi-amplitude of probabilities, i.e. a phase-space wave function associated to the Wigner function \cite{oliveira,size6}. In order to consider the phase space structure of a quantum system in a torus, this formalism in phase space is  considered with TFD.

There are three different approaches to consider finite temperature in quantum field theory: (i) imaginary time formalism~\cite{Matsubara}; (ii) closed time path formalism \cite{Sch, Keld}; and (iii) the TFD formalism \cite{ Umezawa1, Umezawa2, Umezawa22, Khanna1, Khanna2}. The TFD formalism is used here. The statistical average of an arbitrary operator is interpreted as an expectation value in a thermal vacuum. Two  elements to construct the thermal state are necessary: doubling of the Hilbert space: a) original space and b) dual space with Bogoliubov transformations. The doubling is defined by dual conjugation rules. The effects of temperature are introduced by a rotation between the original and the dual Hilbert space. This structure represents a field theory on a torus in the time coordinate, where the length of the circumference is the temperature. The formalism is generalised to describe the quantum field in hyper (spatial) torus. Here it is used for the Schr\"odinger field in phase space.

The paper is organised as follows. In Section II, in order to fix the notation, an outline of the Galilean covariance in phase space is presented considering Bosons and Fermions. In Section III,  the TFD formalism in phase phase is derived. In Section IV,  the size-effect at finite temperature is studied for Bosons; while Fermions are analysed in Section V. In Section VI, some concluding remarks are presented.

\section{Symplectic Galilean Covariance} \label{gc}
In this section we review some aspects of the Galilei covariance in phase space. Then we fix the notation focusing on aspects that will be useful for next sections.
\subsection{Galilean Covariance}
The Galilean transformations that describe classical physics are given by
$
\overline{\mathbf{x}} = R\mathbf{x+v}t+\mathbf{a} ;\,\,\,
\overline{t} = t+b  \label{t gal2}
$,
where $R$ is a 3-dimensional Euclidian rotation, $\mathbf{v}$ is the relative velocity defining the Galilean boost, $\mathbf{a}$ stands for
space translation and $b$ for a time translation. They form a group and subgroups that describe temporal, spatial translations, rotations and boost transformations.

In order to construct a theory of quantum particles,  unitary representations of the Galilean group is used. However, it is not possible to achieve this goal with the above transformations \eqref{t gal2}. The Galilean covariance is needed \cite{Levy1967}. The Galilean covariant transformations are obtained by using  the dispersion relation
$
p_{\mu }p^{\mu }=\mathbf{p}^{2}-2mE=0
$,
where $p^{\mu}=(\mathbf{p},p^{4}=E,p^{5}=m)$, $p_{\mu }=\eta_{\mu\nu}p^{\nu }$, such that $ \eta_{\mu \nu} $ is the Galilean metric given by the non-null components~\cite{Santana}: $\eta_{\mu\nu};\,\,  \mu,\nu = 1,2,...,5$, such that $\eta_{\mu\nu}=1$, for $\mu=\nu=1,2,3$; and $\eta_{4,5}=\eta_{5,4}=-1$.

The coordinates in a manifold ${\cal M}$, with a metric $\eta_{\mu\nu}$ transform as $
x^{\mu}\,'=\Lambda^{\mu}\,_{\nu}  x^{\nu}+a^\mu$,
where $a^\mu=({\textbf a}, b,0)$ and $\Lambda^{\mu}\,_\nu $ is the transformation matrix given by
\begin{equation}
\Lambda^{\mu}\,_\nu= \left(
           \begin{array}{ccc}
             R & 0 & -\textbf{v} \\
             -\textbf{v}\cdot R & 1 & \frac{1}{2}\textbf{v}^2 \\
             0 & 0 & 1 \\
           \end{array}
         \right)\,. \label{3.3}
\end{equation}

Using the invariant line element
$
dl^2=\eta_{\mu\nu}dx^{\mu}dx^{\nu}\,,\label{3.6}
$ leads to $ dr^ 2-2dtds = 0 $, resulting in $ds = \frac {{\textbf v}} {2} \cdot{\textbf dr}$, where the fifth coordinate $x^5=s$ is used.  The Lie algebra of the Galilean group is
\begin{equation}
\begin{array}{rcl}\label{gal_alg}
\left[ M_{\mu \nu },M_{\rho \sigma }\right]  & = & i\left( \eta _{\mu \rho
}M_{\nu \sigma }+\eta _{\nu \sigma }M_{\mu \rho }-\eta _{\nu \rho }M_{\mu
\sigma }-\eta _{\mu \sigma }M_{\nu \rho }\right)  \\
{\left[ P_{\mu },M_{\rho \sigma }\right] } & = & -i\left( \eta _{\mu \rho
}P_{\sigma }-\eta _{\mu \sigma }P_{\rho }\right)  \\
{\left[ P_{\mu },P_{\nu }\right] } & = & 0.%
\end{array}%
\end{equation}%
Here the rotation, $M_{ij}\rightarrow \epsilon _{ijk}J_{k}$, the translation, $P_{i}$, the temporal translation, $P_{4}\rightarrow -H$, and the boost, $M_{5i}=-M_{i5}\rightarrow K_{i}$, are defined.
The Casimir invariants for the scalar field is $I_1=p^\mu p_\mu$ and $I_2=p_5$. Then the Schr\"{o}dinger equation is written as
$
\partial^\mu \partial_\mu \psi (x,t)=0
$,
where $\partial_5\psi(x,t)=-im\psi(x,t)$. The Pauli-Schr\"{o}dinger equation in Galilean covariant formalism has the form of a massless Dirac equation, $
\gamma^\mu\partial_\mu\psi(x,t)=0
$. Now results associated with the Wigner formalism are derived.

\subsection{Covariant Symplectic Quantum Mechanics}\label{Symp}

The symplectic quantum mechanics is introduced by a star-operator defined as
\begin{eqnarray}
\widehat{A}^\mu=a^\mu\star &=&a(q,p)\text{exp}\left[\frac{i}{2}\left(\lpdv{q^\mu}\rpdv{p_\mu}-\lpdv{p^\mu}\rpdv{q_\mu}\right)\right]\nonumber
\\
&=&a\left(q^\mu+\frac{i}{2}\frac{\partial}{\partial p_\mu},p^\mu-\frac{i}{2}\frac{\partial}{\partial q_\mu}\right).
\end{eqnarray}
The following operators are obtained
\begin{eqnarray}
\widehat{P}^\mu&=&p^\mu\star=p^\mu-\frac{i}{2}\frac{\partial}{\partial q_\mu},\label{eq__p1}\\
\widehat{Q}^\mu&=&q\star=q^\mu+\frac{i}{2}\frac{\partial}{\partial p_\mu}\label{eq__q1}
\end{eqnarray}
and
\begin{eqnarray}
\widehat{M}_{\nu\sigma}&=&M_{\nu\sigma}\star=\widehat{Q}_\nu\widehat{P}_\sigma-\widehat{Q}_\sigma\widehat{P}_\nu,
\end{eqnarray}
which satisfy the Heisenberg commutation relation $[\widehat{Q}^\mu,\widehat{P}^\nu]=i\eta^{\mu\nu}$. These unitary operators lead to a set of commutation relations
\begin{eqnarray}
\left[\widehat{P}_\mu, \widehat{M}_{\rho\sigma}\right]&=&-i(\eta_{\mu\rho}\widehat{P}^\sigma-\eta_{\mu\sigma}\widehat{P}^\rho),\\
\left[\widehat{P}_\mu, \widehat{P}_{\sigma}\right]&=&0,\\
\left[\widehat{M}_{\mu\nu},\widehat{M}_{\rho\sigma}\right]&=&-i(\eta_{\nu\rho}\widehat{M}_{\mu\sigma}-\eta_{\mu\rho}\widehat{M}_{\nu\sigma}+\eta_{\mu\sigma}\widehat{M}_{\nu\rho}-\eta_{\mu\sigma}\widehat{M}_{\nu\rho}),
\end{eqnarray}
where Eq. \eqref{gal_alg} leads to $P\rightarrow p\star$ and $M\rightarrow M\star$.

Using Eq.~\eqref{eq__p1} and Eq.~\eqref{eq__q1}, the Galilean covariant Schrödinger equation in phase space is
\begin{eqnarray}\label{DKP-22}
     \widehat{P}_{\mu}\widehat{P}^{\mu}\psi(q,p)&=&0,\nonumber\\
      \widehat{P}_{5}\psi(q,p)&=&-m\psi(q,p).
\end{eqnarray}
This leads to the differential equation
\begin{eqnarray}
   \left(p^2-i\text{p}\cdot\nabla-\frac{1}{4}\nabla^2\right)\psi(q,p)=2\left(p_4-\frac{i}{2}\del_t\right)\left(p_5 -\frac{i}{2}\del_5\right)\psi(q,p).
\end{eqnarray}
This equation, and its complex conjugate, is derived from the Lagrangian density,
\begin{eqnarray}
    \mathcal{L}&=&\frac{1}{4}\del^\mu\psi(q,p)\del_\mu\psi^\dagger(q,p)+\frac{i}{2}p^\mu[\psi(q,p)\del_\mu\psi^\dagger(q,p)\nonumber\\
    &-&\psi^\dagger(q,p)\del_\mu\psi(q,p)]+\left[p^\mu p_\mu\right]\psi(q,p)\psi^\dagger(q,p),
\end{eqnarray}
where $\partial^\mu=\frac{\partial}{\partial q_\mu}$. The physical interpretation of the formalism \cite{oliveira, amorim} is associated with the Wigner function, $f_w$, given as $f_w(q,p)=\psi(q,p)\star\psi^\dagger(q,p)
$. This implies that the Wigner function is not a probability density.  The negativity structure may be related to non-classicality.

It is important to observe that using the notion of gauge symmetry it is possible to introduce a gauge field through the notion of covariant derivative, such that
$\partial_\mu \rightarrow D_\mu = \partial_\mu + eA_\mu\star$. Consider 4th-component of $A_\mu$ be the scalar potencial, we have $A_4\star=\phi(x)\star$ (the other components are null). For the linear approximation, simulating a quark-antiquark bound state, we have $\phi(x)\star=\lambda x\star= \lambda( x +(i/2)\partial_x)$. This result will be explored elsewhere.

\subsection{ Symplectic Spin 1/2 Representation}\label{spin_rep}
The Lagrangian density for spin-1/2 field is
\begin{equation}
\mathcal{L}=-\frac{i}{4}\Big((\del_\mu\bar{\psi}(q,p))\gamma^\mu\psi(q,p)-\bar{\psi}(q,p)(\gamma^\mu\del_\mu\psi(q,p))\Big)+\bar{\psi}(q,p)\gamma^\mu p_\mu\psi(q,p),
\end{equation}
where $\bar{\psi}=\zeta\psi^\dagger$, with $\zeta=-\frac{i}{\sqrt{2}}\{\gamma^4+\gamma^5\}=
\left(\begin{array}{cc}
0&-i\\
i&0
\end{array}\right).$

The Galilean covariant Pauli-Schr\"{o}dinger equation is given as
\begin{equation}
     \gamma^\mu\left(p_\mu-\frac{i}{2}\partial_\mu\right)\psi(p,q)=0.\label{eq:DPS}
\end{equation}

The connection to the Wigner function is given by
$f_w(q,p)= \psi(q,p)\star\bar{\psi}(q,p).$
The energy-momentum tensor for the Dirac-like field in phase-space is obtained by using the Noether theorem as
\begin{eqnarray}
T^{\mu\nu}(q,p)=-\frac{i}{4}\Bigg(-\bar{\psi}(q,p)\gamma^\mu\frac{\partial \psi(q,p)}{\partial q_\nu}+\frac{\partial \bar{\psi}(q,p)}{\partial q_\nu}\gamma^\mu\psi(q,p)\Bigg)-\eta^{\mu\nu}\mathcal{L}(q,p).
\end{eqnarray}

The Green function, $G_D(q-q^\prime, p-p^\prime)$, in phase space is obtained from the equation
\begin{eqnarray}
\gamma^\mu\bigg(\frac{i}{2}\partial_\mu-(p_\mu-p^\prime_\mu)\bigg)G_D(q-q^\prime,p-p^\prime)=\delta(q-q^\prime)\delta(p-p^\prime).
\end{eqnarray}
For $k=q-q'$ it leads to
\bea
\gamma^\mu\bigg(\frac{1}{2}k_\mu-(p_\mu-p^\prime_\mu)\bigg)\overline{G}(k,p-p^\prime)=\delta(p-p^\prime)
\eea
with $\overline{G}(k,p-p^\prime)=\frac{1}{(2\pi)^4}\int d^4qe^{ik^\mu(q_\mu-q^\prime_\mu)}G_D(q-q^\prime,p-p^\prime)$.
Then the field propagator is
\bea
\overline{G}(k,p-p^\prime)=\frac{\delta(p-p^\prime)}{\gamma^\mu\bigg[\frac{1}{2}k_\mu-\big(p_\mu-p^\prime_\mu\big)\bigg]}.
\eea
This leads to
\bea
G_D(q-q^\prime,p-p^\prime)&=&\int\frac{d^5k}{(2\pi)^5}e^{-ik^\mu(q_\mu-q^\prime_\mu)}\overline{G}(k,p-p^\prime)\nonumber\\
&=&\int\frac{d^5k}{(2\pi)^5}e^{-ik^\mu(q_\mu-q^\prime_\mu)}\frac{\delta(p-p^\prime)}{\gamma^\mu\bigg[\frac{1}{2}k_\mu-\big(p_\mu-p^\prime_\mu\big)\bigg]}\,,
\eea
which is the Green function for use at finite temperature. In the next section, the propagator is used to consider the theory in a toroidal topology at a finite temperature.

\section{Symplectic TFD}\label{TFD}

The Thermo Field Dynamics (TFD) formalism is introduced for studying fields at finite temperatures and in toroidal topologies~\cite{ Umezawa1, Umezawa2, Umezawa22, Khanna1, Khanna2}. Here the approach is extended to define Wigner function. The thermal average of an observable is considered as the vacuum expectation value in an extended Fock space, i.e.,  ${\cal S}_T={\cal S}\otimes \tilde{\cal S}$, where ${\cal S}$ and $\tilde{\cal S}$ are the  original and tilde space respectively. This defines the Bogoliubov transformation. The relation between the tilde $\widetilde{\cal X}_i$ and non-tilde ${\cal X}_i$ operators is defined as
\bea
({\cal X}_i{\cal X}_j)^\thicksim = \widetilde{{\cal X}_i}\widetilde{{\cal X}_j}, \quad (c{\cal X}_i+{\cal X}_j)^\thicksim = c^*\widetilde{{\cal X}_i}+\widetilde{{\cal X}_j}, \quad\quad ({\cal X}_i^\dagger)^\thicksim = \widetilde{{\cal X}_i}^\dagger, \quad\quad (\widetilde{{\cal X}_i})^\thicksim = -\xi {\cal X}_i\,,
\eea
with $\xi = -1$ for bosons and $\xi = +1$ for fermions. The Bogoliubov transformation that describes the finite temperature effect among variables leads to
\bea
\left( \begin{array}{cc} {\cal X}(k, \alpha)  \\ \widetilde {\cal X}^\dagger(k, \alpha) \end{array} \right)={\cal B}(\alpha)\left( \begin{array}{cc} {\cal X}(k)  \\ \widetilde {\cal X}^\dagger(k) \end{array} \right),
\eea
and the Bogoliubov transformation, ${\cal B}(\alpha)$, is
\bea
{\cal B}(\alpha)=\left( \begin{array}{cc} u(\alpha) & -v(\alpha) \\
\xi v(\alpha) & u(\alpha) \end{array} \right),
\eea
where $u^2(\alpha)+\xi v^2(\alpha)=1$. The $\alpha$ parameter defined as the compactification parameter given by $\alpha=(\alpha_0,\alpha_1,\cdots\alpha_{D-1})$. The temperature effect is described by the choice $\alpha_0\equiv\beta$ and $\alpha_1,\cdots\alpha_{D-1}=0$, where $\beta=\frac{1}{k_BT}$ with $T$ being the temperature and $k_B$ the Boltzmann constant.

The propagator for the scalar field is
\bea
G_0^{(ab)}(q-q',p-p' ;\alpha)=i\langle 0,\widetilde{0}| \tau[\psi^a(q,p;\alpha)\psi^b(q',p';\alpha)]| 0,\widetilde{0}\rangle,
\eea
where,
\bea
\psi(q,p;\alpha)&=&{\cal B}(\alpha)\psi(q,p){\cal B}^{-1}(\alpha).
\eea
Here $a, b=1,2$ and $\tau$ is the time ordering operator. Then the propagator at the thermal vacuum $|0(\alpha)\rangle={\cal B}(\alpha)|0,\widetilde{0}\rangle$ is
\bea
G_0^{(ab)}(q-q',p-p';\alpha)&=&i\langle 0(\alpha)| \tau[\psi^a(q,p)\psi^b(q',p')]| 0(\alpha)\rangle,\nonumber\\
&=&i\int \frac{d^5k}{(2\pi)^5}e^{-ik(q-q')(p-p')}G_0^{(ab)}(k;\alpha),
\eea
where
\bea
G_0^{(ab)}(k;\alpha)={\cal B}^{-1}(\alpha)G_0^{(ab)}(k){\cal B}(\alpha),
\eea
with
\bea
G_0^{(ab)}(k)=\left( \begin{array}{cc} G_0(k) & 0 \\
0 & G^*_0(k) \end{array} \right).
\eea
Then $G_0(k)$ is given as
\bea
G_0(k)=\frac{\delta^5(p-p^\prime)}{\frac{1}{4}k^2-(p^\mu-p'^\mu) k_\mu+(p^\mu-p'^\mu) (p_\mu-p'_\mu)+i\epsilon},
\eea
where $\delta^5(p-p^\prime)$ is the Dirac delta function \cite{abreu2003}.
Then the non-tilde variable is
\bea
G_0^{(11)}(k;\alpha)=G_0(k)+\xi v^2(k;\alpha)[G^*_0(k)-G_0(k)],
\eea
where $v^2(k;\alpha)$ is the generalized Bogoliubov transformation \cite{Khanna:2011gf} given as
\bea
v^2(k;\alpha)=\sum_{s=1}^d\sum_{\lbrace\sigma_s\rbrace}2^{s-1}\sum_{l_{\sigma_1},...,l_{\sigma_s}=1}^\infty(-\eta)^{s+\sum_{r=1}^sl_{\sigma_r}}\,\exp\left[{-\sum_{j=1}^s\alpha_{\sigma_j} l_{\sigma_j} k^{\sigma_j}}\right],\label{BT}
\eea
with $d$ being the number of compactified dimensions, $\eta=1(-1)$ for fermions (bosons) and $\lbrace\sigma_s\rbrace$ denotes the set of all combinations with $s$ elements.
In the next sections, these results are applied to analyze the quark-antiquark bound states described by a Schr\"odinger equation. It is interesting to evaluate the vacuum fluctuation considering that the system is confined. The confinement is described by a toroidal structure, taking the limit of free-quarks, in the linear potential. First the usual situation of the boson field is considered. Later to take into account the calculation for the spin 1/2 quarks.

\section{Non-relativistic Stefan-Boltzmann law and size-effect in phase space: Boson}\label{results}

In this section the Stefan-Boltzmann law and the size-effect at finite temperature are obtained for Schr\"{o}dinger (Klein-Gordon-like)  equation in phase space.

\subsection{Schr\"{o}dinger equation}

The Galilean Lagrangian density for covariant scalar field in phase space is 
\begin{eqnarray}
    \mathcal{L}_\psi(q,p)&=&\frac{1}{4}\del^\mu\psi(q,p)\del_\mu\psi^\dagger(q,p)+\frac{i}{2}p^\mu[\psi(q,p)\del_\mu\psi^\dagger(q,p)\nonumber \\
    &-&\psi^\dagger(q,p)\del_\mu\psi(q,p)]+\left[p^\mu p_\mu\right]\psi(q,p)\psi^\dagger(q,p).
\end{eqnarray}

The energy-momentum tensor for bosons is 
\begin{eqnarray}
T^{\mu\nu}(q,p)_B&=&\frac{\partial \mathcal{L}_\psi(q,p)}{\partial(\partial_\mu \psi(q,p))}\partial^\nu\psi(q,p)-\eta^{\mu\nu}\mathcal{L_\psi}(q,p).
\end{eqnarray}
This leads to
\begin{eqnarray}
T^{\mu\nu}(q,p)_B&=&\frac{1}{4}\del^\mu\psi(q,p)\del^\nu\psi^\dagger(q,p)+\frac{i}{2}p^\mu[\psi(q,p)\del^\nu\psi^\dagger(q,p)-\psi^\dagger(q,p)\del^\nu\psi(q,p)]\nonumber\\
    &-&\eta^{\mu\nu}\bigg[\frac{1}{4}\del^\lambda\psi(q,p)\del_\lambda\psi^\dagger(q,p)+\frac{i}{2}p^\lambda[\psi(q,p)\del_\lambda\psi^\dagger(q,p)-\psi^\dagger(q,p)\del_\lambda\psi(q,p)]\nonumber\\
&+&\left[p^\lambda p_\lambda\right]\psi(q,p)\psi^\dagger(q,p)\bigg].
\end{eqnarray}
It is necessary to re-write the energy-momentum tensor, at different space-time points as
\bea
T^{\mu\nu}(q,p)_B&=&\lim_{(q'^\mu, p'^\mu)\rightarrow (q^\mu, p^\mu)}\tau\Bigg\{\frac{1}{4}\bigg(\frac{\partial\psi'^\dagger(q,p)}{\partial q'_\mu}\frac{\partial\psi(q,p)}{\partial q_\nu}+\frac{\partial\psi'^\dagger(q,p)}{\partial q'_\nu}\frac{\partial\psi(q,p)}{\partial q_\mu}\bigg)\nonumber\\
&+&\frac{i}{2}(p^\mu-p'^\mu)\bigg[\psi(q,p)\frac{\partial\psi'^\dagger(q,p)}{\partial q'_\nu}-\psi'^\dagger(q,p)\frac{\partial\psi(q,p)}{\partial q_\nu}\bigg]-\eta^{\mu\nu}\bigg[\frac{1}{4}\del^\lambda\psi(q,p)\del_\lambda\psi^\dagger(q,p)\nonumber\\
&+&\frac{i}{2}(p^\lambda-p'^\lambda)\bigg[\psi(q,p)\del_\lambda\psi'^\dagger(q,p)-\psi^\dagger(q,p)\del_\lambda\psi(q,p)\bigg]\nonumber\\
&+&(p^\lambda-p'^\lambda) (p_\lambda-p'_\lambda)\psi(q,p)\psi'^\dagger(q,p)\bigg]\Bigg\}\nonumber\\
&=&\lim_{(q'^\mu, p'^\mu)\rightarrow (q^\mu,  p^\mu)}\Gamma^{\mu\nu}_1\tau\{\psi(q,p)\psi^\dagger(q,p)\}
\eea
where $\tau$ is the time ordering operator and
\begin{eqnarray}
\Gamma^{\mu\nu}_1&=& \frac{1}{4}\bigg(\frac{\partial}{\partial q'_\mu}\frac{\partial}{\partial q_\nu}+\frac{\partial}{\partial q'_\nu}\frac{\partial}{\partial q_\mu}\bigg)+\frac{i}{2}(p^\mu-p'^\mu)\bigg[\frac{\partial}{\partial q'_\nu}-\frac{\partial}{\partial q_\nu}\bigg]\nonumber\\
&-&\eta^{\mu\nu}\bigg\{\frac{1}{4}{\del^\lambda}^\prime\del_\lambda+\frac{i}{2}(p^\lambda-p'^\lambda)\bigg[\frac{\partial}{\partial q'^\lambda}-\frac{\partial}{\partial q^\lambda}\bigg]+\left[(p^\lambda-p'^\lambda) (p_\lambda-p'_\lambda)\right]\bigg\}.\label{gamma}
\end{eqnarray}

The vacuum expectation value of the energy-momentum tensor is 
\bea
\langle {T^{\mu \nu}}(q,p)_B\rangle=\lim_{(q'^\mu, p'^\mu)\rightarrow (q^\mu,  p^\mu)}\left\{\, \Gamma^{\mu\nu}_1G_0(q^\mu-q'{^\mu},p^\mu-p'{^\mu})\right\}.
\eea
The Green function propagator in phase space is
\bea
G_0(q^\mu-q'{^\mu},p^\mu-p'{^\mu})&=&\left\langle 0\left|\tau\big[\psi(q,p)\psi'^\dagger(q',p')\big]\right|0\right\rangle,
\nonumber\\
\nonumber\\
&=& \int\frac{d^5k}{(2\pi)^5} \frac{\delta^5(p-p^\prime)e^{ik_\mu(q^\mu-q'^\mu)}}{\frac{1}{4}k^2- k_\mu(p^\mu-p'^\mu)+(p^\mu-p'^\mu)( p_\mu-p'_\mu)}\nonumber\\
\nonumber\\
&=&\frac{\delta^3(\textbf{p}-\textbf{p}')\delta(p_4-p'_4)}{(2\pi)^3}\theta(t-t')e^{-2im(s-s')}\int d^3k\,e^{i\left[{\textbf k}\cdot ({\textbf{x}}-{\textbf{x'}})-4(K^0+p_4)(t-t')\right]},\nonumber
\eea
with $K^0=\frac{1}{2m}(\frac{\textbf{k}}{2}-\textbf{p})^2$. After integration the propagator is
\begin{eqnarray}
G_0(q^\mu-q'{^\mu},p^\mu-p'{^\mu})&=&\delta^3(\textbf{p}-\textbf{p}')\delta(p_4-p'_4)\bigg(\frac{m}{2\pi}\bigg)^{3/2}\frac{\sqrt{i}}{(t-t')^{3/2}}\theta(t-t')\exp\bigg\{\frac{(p_4-p'_4)(t-t')}{m}\bigg\}\nonumber\\
\nonumber\\
&\times& \exp\bigg\{-im(s-s')+\frac{im(\textbf{x}-\textbf{x}')^2+2i(\textbf{p}-\textbf{p}')\cdot(\textbf{x}-\textbf{x}')(t-t')}{2(t-t')}\bigg\}
\end{eqnarray}

The finite energy-momentum tensor for bosons, with the $\alpha$-parameter, is
\bea
{\cal T}^{\mu \nu (ab)}(q,p;\alpha)_B=\lim_{(q'^\mu, p'^\mu)\rightarrow (q^\mu,  p^\mu)}\left\{\, \Gamma^{\mu\nu}_1\overline{G}_0^{ab}(q^\mu- q'^\mu,  p^\mu - p'^\mu;\alpha)\right\},
\eea
where ${\cal T}^{\mu\nu (ab)}(q,p;\alpha)_B=\langle {T}^{\mu\nu(ab)}(q,p;\alpha)_B\rangle-\langle {T}^{\mu\nu(ab)}(q,p)_B\rangle$ and
\bea
\overline{G}_0^{(ab)}(q^\mu- q'^\mu,  p^\mu - p'^\mu;\alpha)=G_0^{(ab)}(q^\mu- q'^\mu,  p^\mu - p'^\mu;\alpha)-G_0^{(ab)}(q^\mu- q'^\mu,  p^\mu - p'^\mu).
\eea

Some applications for the Klein-Gordon-like equation with different $\alpha$-parameters are investigated. It is important to note that an average component of the energy-momentum tensor diverges in the limit $q'^\mu\rightarrow q^\mu$ at zero temperature.

\subsection{Non-relativistic Stefan-Boltzmann Law}

To calculate the Stefan-Boltzmann law at finite temperature, $\beta$, the $\alpha$-parameter has a value of $\alpha=(0,0,0,\beta,0)$. Then the generalized Bogoliubov transformation is
\bea
v^2(\beta)=\sum_{j=1}^{\infty}e^{-\beta kj}\label{BT1}
\eea
and the Green function becomes
\bea
\overline{G}_0^{(ab)}(q^\mu- q'^\mu,  p^\mu - p'^\mu;\beta)=2\sum_{j=1}^{\infty}G_0(q^\mu- q'^\mu-i\beta j n_4,p^\mu - p'^\mu),\label{GF1}
\eea
where $n_4=(0,0,0,1,0)$. Thus, the components of energy-momentum tensor for bosons with $\mu=\nu=4$ is
\bea
{\cal T}^{44 (11)}(q,p,\beta)_B&=&2\lim_{(q'^\mu, p'^\mu)\rightarrow (q^\mu,  p^\mu)}\sum_{j=1}^{\infty} \Gamma^{44}_1 G_0^{(11)}\left(q^\mu- q'^\mu-i\beta j n_4,p^\mu - p'^\mu\right),\label{STL}
\eea
where
\bea
\Gamma^{44}_1=\frac{1}{2}\partial'_5\partial_5+\frac{i}{2}(p_5-p'_5)(\partial'_5-\partial_5).\label{G44}
\eea
This leads to the form
\bea
{\cal T}^{44 (11)}(p,\beta)_B&=&\lim_{p'^\mu\rightarrow  p^\mu}\delta^3(p-p')\delta(p_4-p'_4)\,\sum_{j=1}^{\infty}\left[ \frac{m^{7/2}-2(p_5-p'_5)m^{5/2}}{(2\pi\beta j)^{3/2}}\right]e^{\frac{i\beta j(p_4-p'_4)}{m}}.
\eea
With $p_5=p'_5=m$ this becomes
\bea
{\cal T}^{44 (11)}(p,\beta)_B&=&\lim_{p'^\mu\rightarrow  p^\mu}\delta^3(p-p')\delta(p_4-p'_4)\,\sum_{j=1}^{\infty}\left[ \frac{m^{7/2}}{(2\pi\beta j)^{3/2}}\right]e^{\frac{i\beta j(p_4-p'_4)}{m}}.
\eea
This is the non-relativistic Stefan-Boltzmann law in phase space. It is important to note that, performing integration of the momenta, the non-relativistic Stefan-Boltzmann law in configuration space, i.e. $\mathbb{T}^{44 (11)}(\beta)_B$, is recovered, i.e.,
\bea
\mathbb{T}^{44 (11)}(\beta)_B&=&\left[ \frac{m^{7/2}}{(2\pi\beta )^{3/2}}\right]\zeta\left(\frac{3}{2}\right),
\eea
as calculated in \cite{Our0}.

\subsection{Non-relativistic size-effect at finite temperature}

Taking $\alpha=(0,0,i2d,\beta,0)$ the Bogoliubov transformation is
\bea
v^2(\beta,d)&=&\sum_{j=1}^\infty e^{-\beta kj}+\sum_{l=1}^\infty e^{-i2dkl}+2\sum_{j,l=1}^\infty e^{-\beta kj-i2dkl}.\label{BT33}
\eea
The first two terms are associated with the Stefan-Boltzmann law and the Casimir effect at zero temperature. The Green function of the third term is
\bea
\overline{G}_0(q^\mu- q'^\mu,  p^\mu - p'^\mu;\beta,d)=4\sum_{j,l=1}^\infty G_0\left(q^\mu- q'^\mu-i\beta jn_4-2dln_3, p^\mu - p'^\mu\right)\label{GF3}
\eea
with $n_3=(0,0,1,0,0)$. Then the finite energy-momentum tensor for bosons with $\mu=\nu=3$ becomes
\bea
{\cal T}^{33 (11)}(q,p,\beta, d)_B&=&4\lim_{(q'^\mu, p'^\mu)\rightarrow (q^\mu,  p^\mu)}\sum_{j,l=1}^{\infty} \Gamma^{33}_1 G_0^{(11)}\left(q^\mu- q'^\mu-i\beta jn_4-2dln_3, p^\mu - p'^\mu\right),
\eea
where
\bea
\Gamma^{33}_1&=&\frac{1}{4}\left(\partial'_3\partial_3-\partial'_1\partial_1-\partial'_2\partial_2+\partial'_4\partial_5+\partial'_5\partial_4\right)\nonumber\\
&-&\frac{i}{2}\left[(p^1-p'^1)(\partial'_1-\partial_1)+(p^2-p'^2)(\partial'_2-\partial_2)+(p^4-p'^4)(\partial'_4-\partial_4)+(p^5-p'^5)(\partial'_5-\partial_5)\right]\nonumber\\
&-&(p^\lambda-p'^\lambda) (p_\lambda-p'_\lambda).
\eea

Therefore the non-relativistic size-effect in phase space at finite temperature is 
\bea
{\cal T}^{33 (11)}(p,\beta, d)_B&=&\lim_{p'^\mu\rightarrow p^\mu}\delta^3(p-p')\delta(p_4-p'_4)\sum_{j,l=1}^{\infty}\frac{e^{\frac{-2d^2l^2m}{\beta j}+2dil(p_3-p'_3)+\frac{i\beta j(p_4-p'_4)}{m}}m^{1/2}}{(2\pi)^{3/2}(\beta j)^{7/2}}\nonumber\\
&\times &\Biggl[8(dl)^2m^3-2m\beta j[m-2\beta j(p^\lambda-p'^\lambda)(p_\lambda-p'_\lambda)]+5m(\beta j)^2(p_1-p'_1)^2\nonumber\\
&+& 5m(\beta j)^2(p_2-p'_2)^2+4im^2dl\beta j(p_3-p'_3)-m(\beta j)^2(p_3-p'_3)^2\nonumber\\
&-&(p_4-p'_4)[6m\beta j+2im(\beta j)^2+4i(\beta j)^2]\Biggl].
\eea
It is to be noted that the non-relativistic size-effect in the configuration space, i.e., $\mathbb{T}^{33 (11)}(\beta, d)_B$, as calculated in \cite{Our0}, is recovered performing the integration in the momenta, i.e.,
\bea
\mathbb{T}^{33(11)}(\beta,d)_B&=&2m\left(\frac{m}{2\pi}\right)^{3/2}\sum_{j,l=1}^{\infty}\left[\frac{4md^2l^2}{(\beta j)^{7/2}}-\frac{1}{(\beta j)^{5/2}}\right]e^{-\frac{2md^2l^2}{\beta j}}.
\eea
It is important to note that the total energy momentum tensor in the configuration space is given by three terms, i.e.,  $$\mathbb{T}^{33(11)}(\beta,d)_{B(T)}=\mathbb{T}^{33(11)}(\beta)_B+\mathbb{T}^{33(11)}(d)_B+\mathbb{T}^{33(11)}(\beta,d)_B.$$  In the right-hand size, the first term is the Stefan-Boltzmann law; the second is a divergent term that can be renormalized out, and the third is the size-effect remaining term, which is convergent and goes to zero in the limit of $\beta,d \rightarrow \infty$, consistently. The behavior of the third term as a function of temperature is shown in the Fig. 1. We have considered the toroidal size-effect , $d$ as $1$fm, that is of the order of a hadron. The quark mass we take the effective value $m = 350$ MeV. It is to be note that at $t=0$, the expected result, and the energy-momentum tensor goes to zero. However, there is an unexpected divergence for the high temperature. In order to improve the result, the case of spin 1/2 for the quark is considered in the next section.

\begin{figure}[h]
\includegraphics[scale=1]{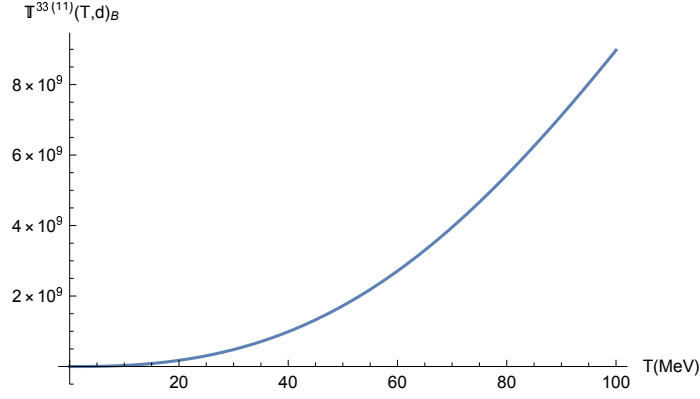}
\caption{Pressure, $\mathbb{T}^{33(11)}(\beta,d)_B$, versus temperature  for $d=1\,\mathrm{fm}=0.005\,\mathrm{MeV}^{-1}$ and $m=350\,\mathrm{MeV}$.}
\end{figure}


\section{Non-relativistic Stefan-Boltzmann law and size-effect in phase space: Fermion}
The Pauli-Schrödinger equation in phase space is described by the Dirac-like Lagrangian density in the Galilean manifold as
\begin{eqnarray}
    \mathcal{L}&=&-\frac{i}{4}\bigg(\partial_\mu\bar{\psi}\gamma^\mu\psi-\bar{\psi}\gamma^\mu\partial_\mu\psi\bigg)-\bar{\psi}\gamma^\mu p_\mu\psi.
\end{eqnarray}
The Galilean Dirac-like propagator is
\bea
G_D(q-q^\prime,p-p^\prime)&=&\int\frac{d^5k}{(2\pi)^5}e^{-ik^\mu(q_\mu-q^\prime_\mu)}\overline{G}(k,p-p^\prime)\nonumber\\
&=&\int\frac{d^5k}{(2\pi)^5}e^{-ik^\mu(q_\mu-q^\prime_\mu)}\frac{\delta(p-p^\prime)}{\gamma^\mu\bigg[\frac{1}{2}k_\mu-\big(p_\mu-p^\prime_\mu\big)\bigg]},
\eea
with some algebra we obtain
\bea
G_D(q-q^\prime,p-p^\prime)&=&2e^{-i(p^\mu-p'^\mu)(q_\mu-q^\prime_\mu)}\delta(p-p^\prime)\int\frac{d^5k}{(2\pi)^5}\frac{e^{-ik^\mu(q_\mu-q^\prime_\mu)}}{\gamma^\mu k_\mu}.
\eea
This leads to the Green function for the Galilean covariant Pauli-Schrödinger equation in phase space,
\begin{eqnarray}
G_D(q-q^\prime,p-p^\prime)=2e^{-i(p^\mu-p'^\mu)(q_\mu-q^\prime_\mu)}i\gamma^\mu\partial_\mu G_0(q^\mu-q'{^\mu},p^\mu-p'{^\mu}),\label{Diracp}
\end{eqnarray}
where
\begin{eqnarray}
G_0(q^\mu-q'{^\mu},p^\mu-p'{^\mu})&=&\delta^3(\textbf{p}-\textbf{p}')\delta(p_4-p'_4)\bigg(\frac{m}{2\pi}\bigg)^{3/2}\frac{\sqrt{i}}{(t-t')^{3/2}}\theta(t-t')\exp\bigg\{\frac{(p_4-p'_4)(t-t')}{m}\bigg\}\nonumber\\
\nonumber\\
&\times& \exp\bigg\{-im(s-s')+\frac{im(\textbf{x}-\textbf{x}')^2+2i(\textbf{p}-\textbf{p}')\cdot(\textbf{x}-\textbf{x}')(t-t')}{2(t-t')}\bigg\}
\end{eqnarray}
is the scalar field propagator in phase space.

The vacuum average of the energy-momentum tensor associated with the Dirac-like (Fermions) Lagrangian is given by
\begin{eqnarray*}
\left\langle T^{\mu\nu(ab)}(q,p, \alpha)_F\right\rangle&=&\lim_{(q'^\mu, p'^\mu)\rightarrow (q^\mu,  p^\mu)} \Gamma^{\mu\nu}_2 2e^{-i(p^\mu-p'^\mu)(q_\mu-q^\prime_\mu)}i\gamma^\alpha\partial_\alpha G_0(q^\mu-q'{^\mu},p^\mu-p'{^\mu};\alpha),
\end{eqnarray*}
where
\begin{eqnarray*}
\Gamma^{\mu\nu}_2=-\frac{i}{4}\Bigg[-\gamma^\mu\frac{\partial}{\partial q_\nu}+\gamma^\mu\frac{\partial}{\partial q'_\nu}-\eta^{\mu\nu}\bigg(\gamma^\lambda\frac{\partial}{\partial q^\lambda}-\gamma^\lambda\frac{\partial}{\partial q'^\lambda}\bigg)\Bigg]-\eta^{\mu\nu}\gamma^\lambda (p_{\lambda}-p'_\lambda).
\end{eqnarray*}
Then the physical energy-momentum tensor for fermions is defined as
\begin{eqnarray}
{\cal T}^{\mu\nu (ab)}(q,p,\alpha)_F&=&\lim_{(q'^\mu, p'^\mu)\rightarrow (q^\mu,  p^\mu)}2i\gamma^\alpha\partial_\alpha\Gamma^{\mu\nu}_2 \bar{G}_D(q^\mu-q'{^\mu},p^\mu-p'{^\mu};\alpha),
\end{eqnarray}
with ${\cal T}^{\mu\nu (ab)}(q,p;\alpha)_F=\langle {T}^{\mu\nu(ab)}(q,p;\alpha)_F\rangle-\langle {T}^{\mu\nu(ab)}(q,p)_F\rangle$ and $\bar{G}_D(q^\mu-q'{^\mu},p^\mu-p'{^\mu};\alpha)=G_D(q^\mu-q'{^\mu},p^\mu-p'{^\mu};\alpha)-G_D(q^\mu-q'{^\mu},p^\mu-p'{^\mu})$. Here a renormalization procedure has been used. Now results at finite temperature are considered.

\subsection{Non-relativistic Stefan-Boltzmann law }

For the study of the Stefan-Boltzmann law in phase space the parameter $\alpha$ is chosen as $\alpha=(0, 0, 0, \beta, 0)$. The Bogoliubov transformation is
\bea
v^2(k,\beta)=\sum_{j=1}^\infty (-1)^{j+1}\,e^{-\beta k j}\,.
\eea
The Green function for the Dirac-like (Fermions) field in phase space is
\bea
\overline{G}_D^{(ab)}(q- q',  p - p';\beta)&=&\sum_{j=1}^\infty (-1)^{j+1}\bigl[G^*_D(q'-q+i\beta j n_4, p'-p)\nonumber\\
&-&G_D(q-q'-i\beta j n_4,p-p')\bigl].
\eea

By taking $\mu=\nu=4$, the energy-momentum tensor for fermions is
\bea
{\cal T}^{44 (11)}(q,p,\beta)_F&=&\lim_{(q',p')\rightarrow (q,p)}\sum_{j=1}^\infty (-1)^{j+1}\,i\gamma^\alpha\partial_\alpha\Gamma^{44}_2\bigl[G^*_D(q'-q+i\beta j n_4, p'-p)\nonumber\\
&-&G_D(q-q'-i\beta j n_4,p-p')\bigl],
\eea
where
\bea
\Gamma^{44}_2=-\frac{i\gamma^4}{4}\left(\partial_5-\partial'_5\right).
\eea
Then the energy-momentum tensor in phase space becomes
\bea
{\cal T}^{44 (11)}(p,\beta)_F&=&\lim_{p'\rightarrow p}\sum_{j=1}^\infty (-1)^{j+1}\delta^3(\textbf{p}-\textbf{p}')\delta(p_4-p'_4)e^{\frac{-i\beta j(p_4-p'_4)}{m}}\left(\frac{m}{2\pi}\right)^{3/2}\frac{m^2}{(\beta j)^{3/2}},
\eea
where the Clifford algebra $\{\gamma^\mu, \gamma^\nu\}=2\eta^{\mu\nu}$ has been used.
This is the non-relativistic Stefan-Boltzmann law associated with the Pauli-Schr\"{o}dinger equation in the phase space.

In order to recover this result in the configuration space, i.e. $\mathbb{T}^{44 (11)}(\beta)_F$, the integration in the momenta is performed. Then we have
\bea
\mathbb{T}^{44 (11)}(\beta)_F&=&\sum_{j=1}^\infty \frac{(-1)^{j+1}}{j^{3/2}}m^2\left(\frac{m}{2\pi\beta}\right)^{3/2}\nonumber\\
&=&m^2\left(\frac{m}{2\pi\beta}\right)^{3/2}\left[\frac{\sqrt{2}}{2}-1\right]\zeta\left(\frac{3}{2}\right).
\eea

Next we analyse the size-effect.

\subsubsection{Size-effect at finite temperature}

For $\alpha=(0,0,i2d,\beta,0)$, the generalized Bogoliubov transformation becomes
\bea
v^2(\beta,d)&=&\sum_{j=1}^\infty (-1)^{j+1}e^{-\beta k^0j}+\sum_{l=1}^\infty(-1)^{l+1} e^{-i2dk^3l}+2\sum_{j,l=1}^\infty (-1)^{j+l}e^{-\beta k^0j-i2dk^3l}.\label{BT3}
\eea
The first two terms of this expression corresponds to the Stefan-Boltzmann term and the Casimir effect at $T = 0$, respectively. The third term is associated with the size-effect analyzed for $T$ not equal to $0$. It leads to
\bea
{\cal T}^{33 (11)}(q,p,\beta, d)_F&=& \lim_{q'\rightarrow q}\sum_{j,l=1}^\infty\,(-1)^{j+l}i\gamma^\alpha\partial_\alpha\Gamma^{33}_2[G^*_D(q'-q+i\beta j n_4+2dl n_3, p'-p)\nonumber\\
&-&G_D(q-q'-i\beta j n_4-2dl n_3,p-p')\bigl],\label{EMT}
\eea
with
\bea
\Gamma^{33}_2=-\frac{i}{4}\left[-\gamma^3\left(\partial_3-\partial'_3\right)-\gamma^{\lambda}\left(\partial_{\lambda}-\partial'_{\lambda}\right)\right]-\gamma^\lambda(p_\lambda-p'_\lambda).\label{G33}
\eea
Then the non-relativistic size-effect at finite temperature in phase space is
\bea
{\cal T}^{33 (11)}(p,\beta, d)_F&=&\lim_{p'\rightarrow p}\delta^3(\textbf{p}-\textbf{p}')\delta(p_4-p'_4)\sum_{j,l=1}^\infty\,(-1)^{j+l}\frac{e^{-\frac{2md^2l^2}{\beta j}+2idl(p_3-p'_3)-\frac{i\beta j}{m}(p_4-p'_4)}m^{1/2}}{(2\pi)^{3/2}(\beta j)^{7/2}}\nonumber\\
&\times&\Biggl\{4m^3(dl)^2-m^2\beta j+m(\beta j)^2[(p_1-p_1')^2+(p_2-p_2')^2]-4im^2dl\beta j(p_3-p_3')\nonumber\\
&+&2im(\beta j)^2(p_4-p_4')-2m^2(\beta j)^2(p_4-p_4')\Biggl\}.
\eea
This result implies an effect of finite temperature and space compactification. In addition, the usual effect in configuration space, $\mathbb{T}^{33 (11)}(\beta, d)_F$, is obtained after the momentum integration, i.e.,
\bea
\mathbb{T}^{33 (11)}(\beta, d)_F&=&m\left(\frac{m}{2\pi\beta}\right)^{3/2}\sum_{j,l=1}^\infty\,\frac{(-1)^{j+l}}{j^{3/2}}\left[\frac{4md^2l^2-\beta j}{(\beta j)^2}\right]e^{-\frac{2md^2l^2}{\beta j}}.
\eea

As in the case of bosons, the total energy-momentum tensor for fermions in the configuration space is given by three terms, i.e.,
\bea
\mathbb{T}^{33(11)}(\beta,d)_{F(T)}=\mathbb{T}^{33(11)}(\beta)_F+\mathbb{T}^{33(11)}(d)_F+\mathbb{T}^{33(11)}(\beta,d)_F.\label{total}
\eea
In the right-hand size, the first term is the Stefan-Boltzmann law; the second is a divergent term that can be renormalized out, and the third is the size-efect remaining term for this fermion system. This term goes to zero in the limit of $\beta,d \rightarrow \infty$, consistently. The first term, the Stefan-Boltzmann law, in phase space is given as
\bea
{\cal T}^{33(11)}(p,\beta)_F&=&\lim_{p'\rightarrow p}\delta^3(\textbf{p}-\textbf{p}')\delta(p_4-p'_4)\sum_{j,l=1}^\infty\,(-1)^{j+l}\,\frac{\exp\left[-\frac{i\beta j}{m}(p_4-p'_4)\right]m^{1/2}}{(2\pi)^{3/2}(\beta j)^{5/2}}\\
&\times&\Bigl[-m^2+m\beta j(p_1-p_1')^2+m\beta j(p_2-p_2')^2+2im\beta j(p_4-p_4')-2im^2\beta j(p_4-p_4')\Bigl].\nonumber
\eea
Then in the configuration space the energy-momentum tensor becomes
\bea
\mathbb{T}^{33(11)}(\beta)_F=\frac{m^{5/2}}{(2\pi)^{3/2}\,\beta^{5/2}}\left(1-\frac{\sqrt{2}}{4}\right)\zeta\left(\frac{5}{2}\right).
\eea

In order to understand the third term in Eq. (\ref{total}), that is, the size-effect at finite temperature, its behavior is shown in Fig. 2 and Fig. 3. It is important to observe that for $T\rightarrow \infty$, this size-effect term goes to zero, as it would be expected. Increasing the temperature from $T=0$, the contribution of this size-effect term is negative, diminishing the Stefan-Boltzmann effect for the total energy momentum tensor. This term reaches a minimum and starts to grow. The size-effect contribution is zero at $\simeq40$ Mev; such that the Stefan-Boltzmann Boltzmann term provides the unique contribution for the energy-momentum tensor of the (free) quark model. There is an oscillation, which is essentially  due to the fermion nature of the quark (compare with the case of boson). This oscillation, although having no special physical meaning, on the other hand, leads to the correct result of high temperature, as aforementioned.
\begin{figure}[h]
\includegraphics[scale=1]{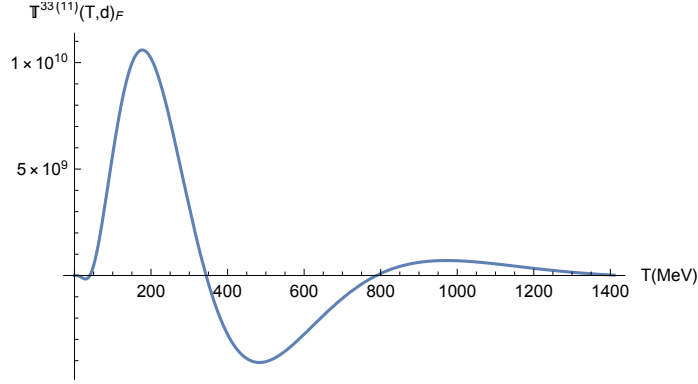}
\caption{Pressure ($\mathbb{T}^{33(11)}(\beta,d)_F$) versus temperature for $d= 1\mathrm{fm}=0.005\,\mathrm{MeV}^{-1}$ and $m=350\,\mathrm{MeV}$. }
\end{figure}

\begin{figure}[h]
\includegraphics[scale=1]{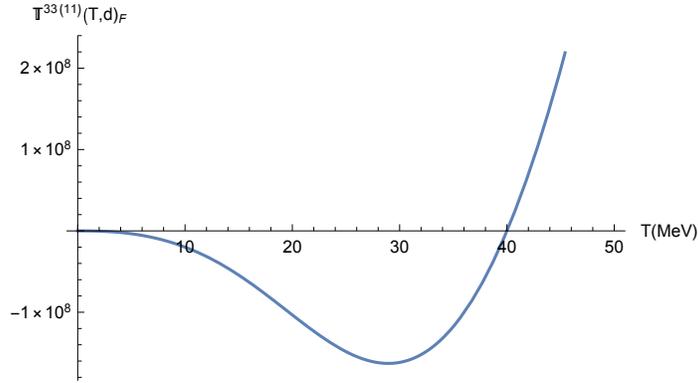}
\caption{This shows only a small part of Fig. 1. This exhibits that the first temperature for which $\mathbb{T}^{33 (11)}(\beta, d)_F\rightarrow 0$ is $T\sim 40\, \mathrm{MeV}$.}
\end{figure}

It is interesting to note that bosons and fermions behave differently when subjected to the thermalization process.  For bosons, there is a constant pressure increase with temperature, so that an experimental limit can stop a divergence.  For fermions, the pressure tends to zero due to the Pauli exclusion principle when the temperature becomes excessively high. When analyzing fermionic behavior, one must ask whether the temperature at which the pressure vanishes first depends on the mass of the quark, since the variety of such particles is notorious.  Thus, in Fig. 4 it can be seen that $\mathbb{T}^{33 (11)}(\beta, d)_F\rightarrow 0$ for $T\sim 140\, \mathrm{MeV}$, for a mass of 100 $\mathrm{MeV}$. In Fig. 5, we have the following behavior for a mass of 750 $\mathrm{MeV}$, $\mathbb{T}^{33 (11)}(\beta, d)_F\rightarrow 0$ for $T\sim 19\, \mathrm{MeV}$. Thus, we conclude that there is a correlation between this phenomenon and the mass of the  quark used, probably representing the confined state of the quark.

\begin{figure}[h]
\includegraphics[scale=1]{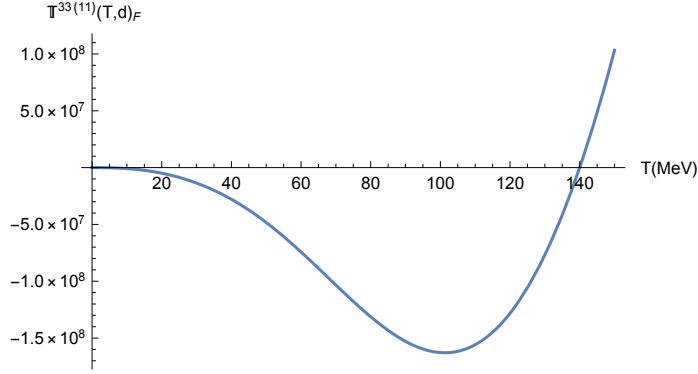}
\caption{Pressure ($\mathbb{T}^{33(11)}(\beta,d)_F$) versus temperature for $d= 1\mathrm{fm}=0.005\,\mathrm{MeV}^{-1}$ and $m=100\,\mathrm{MeV}$.}
\end{figure}

\begin{figure}[h]
\includegraphics[scale=1]{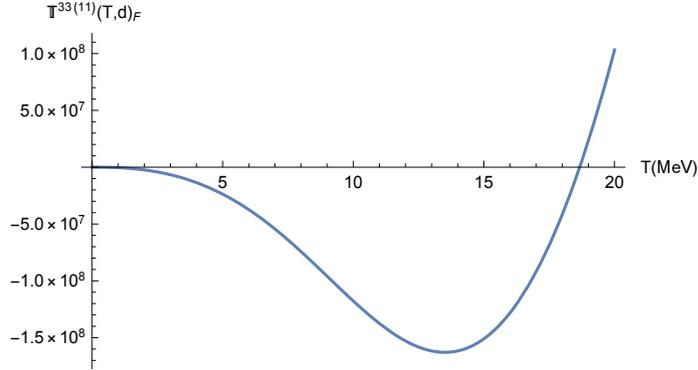}
\caption{Pressure ($\mathbb{T}^{33(11)}(\beta,d)_F$) versus temperature for $d= 1\mathrm{fm}=0.005\,\mathrm{MeV}^{-1}$ and $m=750\,\mathrm{MeV}$.}
\end{figure}

\section{Conclusion} \label{con}

Here three aspects of a quark-antiquark effective model are analysed: (i) the spin structure of the quark; (ii) the non-classicality of such states; (iii) the size-effect at finite temperature. In order to carry out such an analysis, the symplectic generalization of thermo-field dynamics is used.  In order to take into consideration the spin for the quark, a symplectic version of Pauli-Schr\"odinger equation is used.  The Galilean symmetry is assured along the calculations by writing the theory in the light-cone of a   $(4+1)$-de Sitter space.

The energy-momentum tensor associated with each field, Bosons and Fermions, is calculated. Considering the case of bosons, the  size-effect term remains and goes to zero at zero-temperature, as it would be expected. However, the size-effect for high temperature for a fixed compactification lenght remains growing. This is an unexpected result and has motivated an analysis of a more realist model, by considering the spin of a quark. In this case, by increasing the temperature, from $T=0$, the size-effect contribution to the energy momentum tensor is negative, diminishing the Stefan-Boltzmann effect. There is a minimum, and at $T\simeq 40$ Mev the size-effect term of the energy momentum tensor is zero. In this case, the toroidal size-effect, $d$, as the hadron diameter, $d\simeq 1$fm, and the quark mass as $350$ MeV has been considered. In other words,  the size-effect under this condition is no longer important, and the energy-momentum tensor is described only by the Stefan-Boltzmann law, for the (free) fermions. It is important to observe that such a temperature is in order of the estimated temperature for the chiral symmetry breaking, that gives rise to quark-gluon plasma. The connection of these two different results are not simple, but it demands more investigations. Finally, at high temperature, the size-effect term goes to zero, differently from the case of bosons. These results indicate that the spin is a crucial element for the quark-anti-quark effective model.

\section*{Acknowledgments}

This work by A. F. S. is supported by CNPq projects 430194/2018-8 and 313400/2020-2.

\end{document}